# Tuning Catalytic Activity and Selectivity in Photocatalysis on Dielectric Cuprous Oxide Particles

*Ravi Teja Addanki Tirumala[⊥], Sundaram Bhardwaj Ramakrishnan[⊥], Farshid Mohammadparast[⊥], Marimuthu Andiappan\*, [⊥]*

**Affiliations:**

[⊥] School of Chemical Engineering, Oklahoma State University, Stillwater, OK, USA.

\* **Corresponding Author,** Email: mari.andiappan@okstate.edu

**ABSTRACT**

Dye degradation has been for more than forty years in the scientific community. All these studies have primarily focused on breaking various dyes using catalysts driven by either light or heat. Most studies started to focus on metal-oxides after the discovery of water-splitting by $TiO_2$. Among the many catalysts used plasmonic metal nanostructures have been explored significantly in recent times due to their special property called localized surface plasmon resonances (LSPR). However, facing multiple problems of heat losses and instability, people started to focus on dielectric medium-to-high refractive indexed materials for photonic applications. Most of these dielectric materials have been studied from a physics point of view and less from chemistry. In this work, we have focused on how these materials can be used for tuning selectivity through wavelength-dependent studies by performing methylene blue (MB) dye degradation.

**MAIN TEXT**

The field of plasmonics has grown significantly since the coining of the term by Mark L. Brongersma in early 2000. Research and development in plasmonics have grown ever since then exponentially both from the physics and chemistrstandpointnt. The metals gold, silver, aluminum, and copper (Au, Ag, Al, and Cu) fall into this category. These materials have a special property called *localized surface plasmon resonance* (LSPR) which they exhibit when electromagneticectromagentic radiations, namely photons. At resonant frequencies, there is a significant enhancement in the light-matter interaction resulting in high electric fields. The elevated fields generate a large number of charge carriers (electron-hole pairs) which can induce chemical transformations either through localized heating or transfer of charges to the adsorbate on the nanostructure surface. The excitement and growth surrounding plasmonics stem from the ability to use the excitation of energetic charge carriers to drive surface chemistry.[1] These demonstrations show that visible-light-driven chemical transformations on plasmonic metal nanostructures (PMNs) have led to the emergence of a new field in heterogeneous catalysis known as plasmonic photocatalysis. Generally, nanoparticles are an important part of the heterogeneous catalysis being in various chemical reactions such as dehydrogenation, partial oxidations, reduction reactions, ammonia synthesis, hydrocarbon reforming etc.[2–16] However, these plasmonic materials suffer from losses arising from heat and the incompatibility in scaling up on the lines of conventional metal-oxide-semiconductor fabrication. Recent reports have shown a new class of dielectric, medium-high refractive indexed, metal-oxides are playing in important role in nanophotonics. When light interaction in these materials induces *Dielectric-EnhancedMie resonances*, which has reduced heat losses, and enhancing both electric and magnetic near-fields of comaprable strength. While in plasmonics there is only strong eletric fields emhancement. The new class of high indexed dielectric materials exhibit same features as plasmonics like enhanced scattering, nanoantennas, magneticsm and meta-materials making them superior compared to their lossy plasmonic couterparts. So far, these dielectric materials have been studied and reported from physics point of view, they have not studied from the chemistry point of view especically in the field catalysis. Mohammadparast et al. has shown in his recent work that these high-indexed-dielectric materials exhibit strong scattering property acting as nanoantennas by concentrating and directing light.[17] The metal oxide used in the work is $Cu_2O$ particles. The work also shows through finite-difference-time-domain simulation (FDTD) of strong enhancemnt of electric and magnetic fields. In this work we show how Cu₂O particles can be applied to photocatalysis that focuses not only on conversion, but also yield and selectivity. Also, in addition to improving selectivity, we have demostarted through conrete results that, we can tune selectivity through degradation of methylele blue (MB) dye using various LED light sources in the visible-light region. In the process, we narrowed our focus to Cu2O (bandgap ~ 2.1 eV). Compared to plasmonic metals which have multiple modes of electric field excitation due to LSPR, the dielectric Cu2O exhibits multiple-modes of electric and magnetic enhancement upon light excitation. In this study, we are shown how we can selectively degrade methylene blue (MB) peaks through Mie resonance mediated photocatalysis using Cu2O as a catalyst. Most literature studies have shown that catalysts are made from metals, semiconductors, or their hybrids to break MB in the photocatalytic process using either visible-light or ultra-violet. In addition, very little literature work has shown the coupling of dye excitons and Mie resonances in dielectric semiconductors. Here, we are showing the MB degradation using different light sources such as green (510-530 nm), blue (450-460 nm), red (630-650 nm), and amber (585-595 nm) and validate how different light sources affect the outcome of the process differently. As discussed above the catalyst used, is Cu₂O particles. These particles exhibit strong Mie resonance peaks over a broad range of wavelength and act as dielectric nanoantennas or nanoresonators by directing a significant portion of light into their localized neighborhood at nanoscale. Predominantly in MB dye degradation researchers have just focused on primary peak reduction, while in this work we show how can selectively focus different peaks by varying the different wavelengths in the visible-light region. We all know that breaking can happen in multiple pathways depending on the type of catalyst, light source, solvent, and dye itself. The most commonly occurring ones are complete mineralization of dye or dye sensitization followed by dye degradation. All the probe reactions were performed in dimethylformamide (DMF).

This work reports interesting results about how different peaks of MB can be degraded by tuning multiple lights in the visible region. Herein we show some promising new ideas about using medium-refractive indexed semiconductor material being used in achieving this control over breaking down MB peaks.

The probe reactions was carried out in DMF solvent. To the distinguish role of light from heat, reactions in the absence of light were carried out at various temperatures ranging 20-40 ˚C. The absorption spectra MB degradation from samples taken at different time intervals (see, Figure 1a).

The main MB peak (monomer) at 663 nm (see, Figure 3b), decreases with time, in parallel with the increase in trimer peak (570 nm).

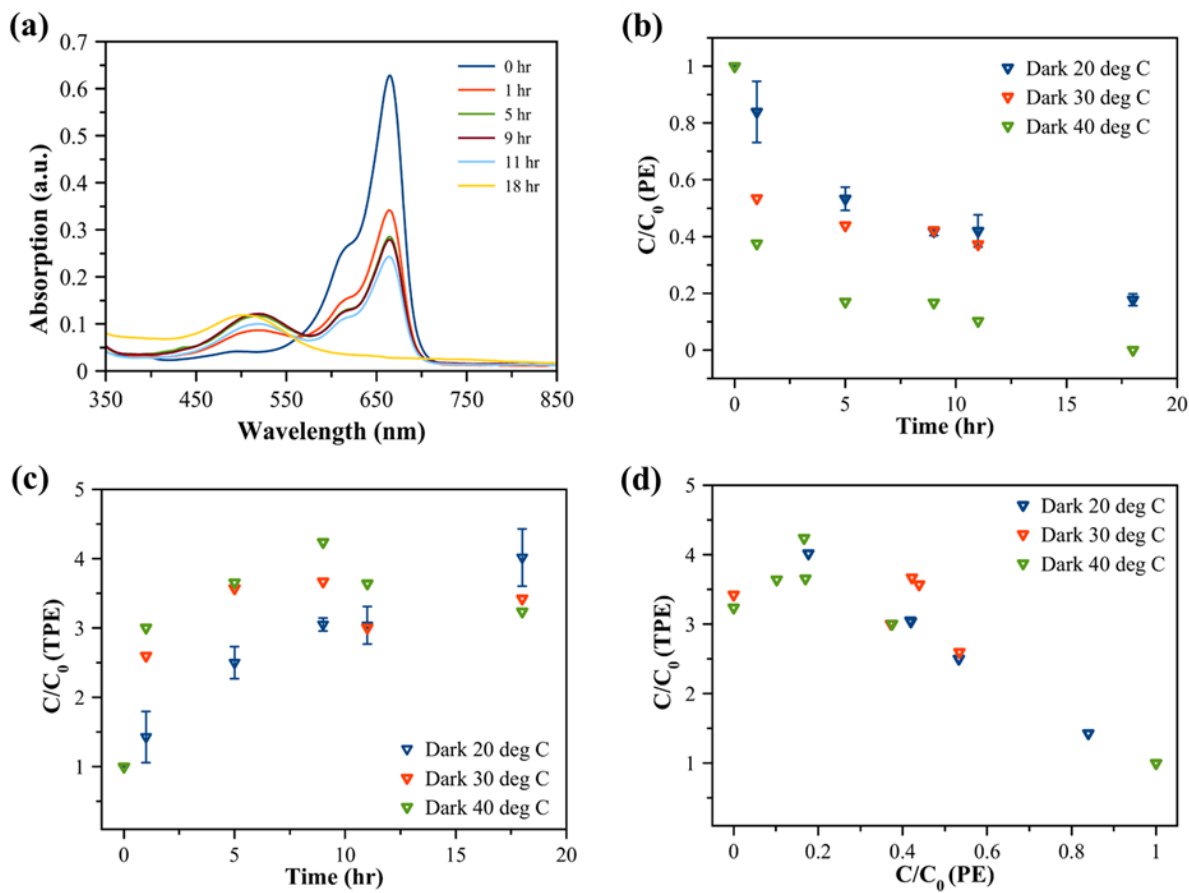

**Figure 1.** Methylene blue (MB) Degradation observed under dark conditions at different temperatures (20-40 ˚C), **(a)** Absorption spectra of degradation of methylene blue at 30 ˚C and dark conditions. **(b)** Degradation of primary absorption peak of MB showing comparison of degradation. **(c)** Increase in tertiary species with time. **(d)** Selectivity comparison showing formation of tertiary absorption peak under various conditions.

It can be seen that in dark conditions, the monomer peak breaks down faster as we raise the temperature from 20 to 40 ˚C (see, Figure 1b). The plot shows the measure conversion of MB into its components such as $CO_2$, $NH_3$, etc. The increase in tertiary peak (trimer) represents the measure of selectivity (see, Figure 1c). We can see that the trimer peak also increases with the temperature rise. This suggests that the rate of disappearance of reactant (monomer) and appearance of the product (trimer) happens at a similar rate. The important observation that we make is at a given conversion the selectivity remains constant for different temperatures (see, Figure 1d). The temperature profiles of the system, ambient, and reactor under these dark and heating conditions suggest that there is consistency in maintaining temperature (see, Figure S1a-c). Now, as we move towards understanding the role of light from temperature we see the extent of degradation using different light sources (see, Figure 2). The green light (see, Figure 2a) shows the least while red shows the most (see, Figure 2b). The separation in the rate of MB breaking process is due to the reaction pathway it follows. These light sources have been benchmarked against dark and heating conditions to show that heating is not responsible for this process, it's mainly driven by light. All these light experiments were performed under constant intensity conditions. The red light has a strong overlap with the monomer peak (663 nm), while the green light has the least (see, Figure 3b). The red light causes dye sensitization followed by complete mineralization resulting in the faster breakdown of the dye, along with strong growth of the trimer peak. On the contrary, if we see the interaction between green light and MB, the monomer remains pretty strong even after 11 hours, but there is no formation of trimer peak(see, Figure 2b). This is due fact that the trimer (570 nm) has a strong overlap with green light. As a result, the trimer that forms immediately disappears due to dye sensitization of the trimer resulting in mineralization forming other products. While in the case of amber, the overlap with MB is in between red and green. In the case of red and amber the overlap triggers the transfer of electrons from the valence band (VB) to CB of the dye, which inturn transfers this electron to the CB of the semiconductor. The pathway for photodegradation is the formation of oxygen radicals that attack the dye resulting in complete mineralization. The resulting rate of MB breakdown in amber stays within red and green. The plot of measure of conversion and measure of selectivity shows that at a given conversion selectivity is different when placed under different light sources (see, Figure 3a). This is due to the fact various light sources overlap with MB peaks over the wider region making different selectivity outcomes . On comparing dark conditions at higher temperature (30 ˚C), where the outcome is slightly faster compared to room temperature. This validates the fact that light sources such as red and amber raise the temperature through light intensity up to ~30 ˚C (see, Figure S2a-b). While comparing the dark at 30 ˚C with green (see, Figure S2c) we can see that it overlaps suggesting that despite rising in reactor temperature we don't see monomer degradation suggesting that these reactions are purely light-driven.

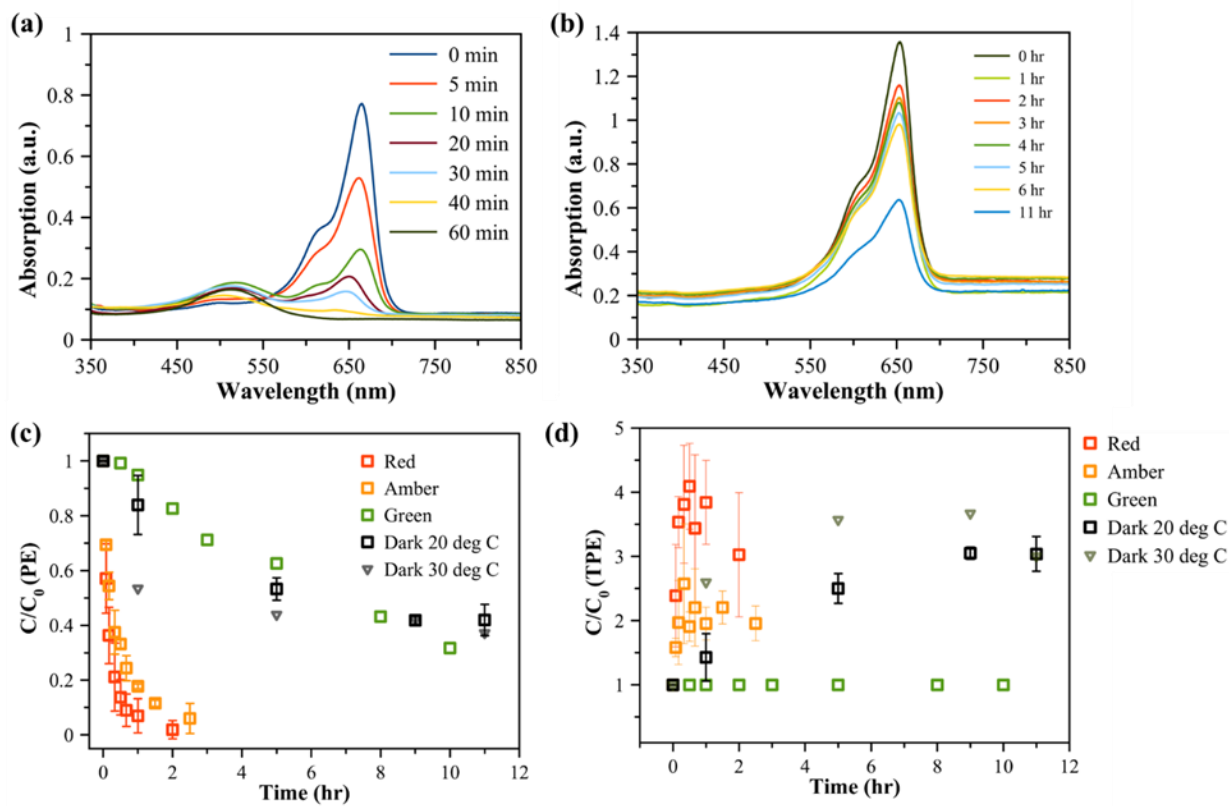

**Figure 2.** Wavelength dependent studies showing Methylene blue (MB) Degradation observed under illumination of different wavelengths and comparison with dark experiments **(a, b)** Absorption spectra of degradation of methylene blue under illumination of red and green light illumination. **(c)** Tuning Selectivity comparison showing formation of tertiary absorption peak of MB under various conditions showing extent of mineralization to tertiary species formed. **(d)** Increase in tertiary species with time is observed except under green light illumination.

The results of blue and UV light are also expected to be on the lines of green or slightly below. This is arising from the fact the both these lights have no overlap with monomer peak, while they have slight overlap with the trimer peak (see, Figure 3b).

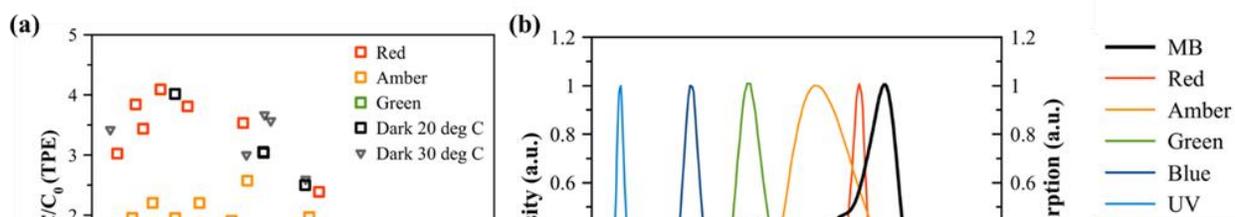

**Figure 3. (a)** Tuning Selectivity comparison showing formation of tertiary absorption peak of MB under various conditions showing extent of mineralization to tertiary species formed **(b)** Exposure standards of different light illumination used in dye degradation experiments. Ref:https://www.luzchem.com/ExposureStandards.php

**Conclusion**

In this study, we show how we can selectively tune the outcomes of breaking down different peaks of MB dye using MPCs and light of appropriate wavelength. Also, giving a brief information about the reaction pathway facilitating the same. This very idea gives strong possibility about selectively degrading various other dye molecules like methyl orange, rhodomine B and other pollutants. The role of light in driving the reaction is well establised from heat driven through multiple base experiments. The notion of selective tuning can be expanded to other areas of photochemistry

# Tuning Catalytic Activity and Selectivity in Photocatalysis on Dielectric Cuprous Oxide Particles

*Ravi Teja Addanki Tirumala[⊥], Sundaram Bhardwaj Ramakrishnan[⊥], Farshid Mohammadparast[⊥], Marimuthu Andiappan\*, [⊥]*

**Affiliations:**

[⊥] School of Chemical Engineering, Oklahoma State University, Stillwater, OK, USA.

\* **Corresponding Author,** Email: mari.andiappan@okstate.edu

## I. Cu₂O nanoparticles Preparation and Characterization

Cu$_2$O particles were prepared using the chemical reduction technique. This technique uses 50 mL DI water, 100 mg PEG (2000) as a surfactant, 100 mg of precursor (CuCl2), 1.5 mL of 3 M NaOH solution, and 100 mg glucose as reducing agents, respectively. 100 mg of PEG (2000) and 100 mg of CuCl2 are added to the 100 mL round bottom flask. Add 50 mL of DI water to the mixture and stirred at 900 rpm. Allow the synthesis medium to stir for 15 minutes and add 1.5 mL of 3M NaOH solution and increase the temperature to 50 ˚C. It is observed that the solution turns blue after the addition of NaOH solution. Allow the synthesis mixture to stir for an hour, after which 100 mg of glucose is added as a reducing agent. After 40 minutes the solution turns crimson, and orange in color. The synthesis mixture is removed and washed (centrifuge and sonicate) with acetone to obtain 47 mg of Cu$_2$O particles.

## II. Methylene Blue Degradation Reaction Conditions and Characterization
## Dark Conditions

Cu$_2$O particles were synthesized using the chemical reduction technique as discussed above, were suspended in dimethylformamide (DMF, 4 mL) before starting the reaction (5.8 mg). The suspended NPs were added to a quartz test tube and sonicated for 2-3 minutes for obtaining a homogenous mixture. 150 uL of 0.01 M methylene blue was added to the mixture as a reactant and stirred at 1150 rpm. The quartz tube was wrapped in aluminum foil/opaque foil from the outside to make sure ambient light is not affecting the reaction mixture. Sampling to obtain absorption spectra as shown in Figure 1d, 100 uL of the reaction mixture in 4 mL of DMF. Car was taken that the same solvent was used across the experimental conditions. The temperature was recorded using a thermocouple before each sample.

## Light Conditions

5.8 mg of catalyst (Cu2O NPS) is suspended in 4 mL DMF and 150 uL 0.01 M MB is added to the concoction similar to dark conditions. The quartz tube is carefully suspended using support into the Luzchem reactor as shown in Figure and the mixture was stirred at 1150 rpm. The luzchem reactor needs to be arranged with bulbs of the desired wavelength before the start of the reaction. 100 uL of the reaction mixture in 4 mL of DMF, sampling was used to track the dye degradation. Light intensity in mW/cm2 is measured using Intellpro.

## Light Intensity Measurements and Constant Intensity Measurements

Measuring Light Intensity was done using Intellpro Intensity meter in Lux and converted to mW/cm2 using a calibration factor specific to wavelength provided. As shown in photocatalytic reactor as shown the Figure (SI) light intensities were measured placing all 20 bulbs of different wavelength in UV-Visible range. Green Light is used as reference for measuring constant light intensity across the spectrum by varying the number of bulbs to obtain constant light intensity conditions irrespective of the wavelength.

## III. Characterizations used:

UV-Vis spectra were obtained using an Agilent Cary 60 Spectrophotometer. Monitoring the degradation of methylene blue in the reaction mixture: For the UV-Vis extinction spectra measurements of the reaction mixture, an aliquot of 100 μL was taken from the reaction mixture and diluted into 4 mL of dimethylformamide(DMF) (100% - Fisher Scientific Cat No. AC423640250) and sonicated for a minute for good homogenization of the mixture which was then used for UV-Vis spectroscopic measurements. XRD patterns were acquired using a Philips X-ray diffractometer (Phillips PW 3710 MPD, PW2233/20 X-Ray tube, Copper tube detector – wavelength - 1.5418 Angstroms), operating at 45 KW, 40 mA. The catalyst weight load of 9.3 % (wt%) in silica was used to perform XRD characterization of Cu2O NPs. The SEM images were acquired using an FEI Quanta 600 F.

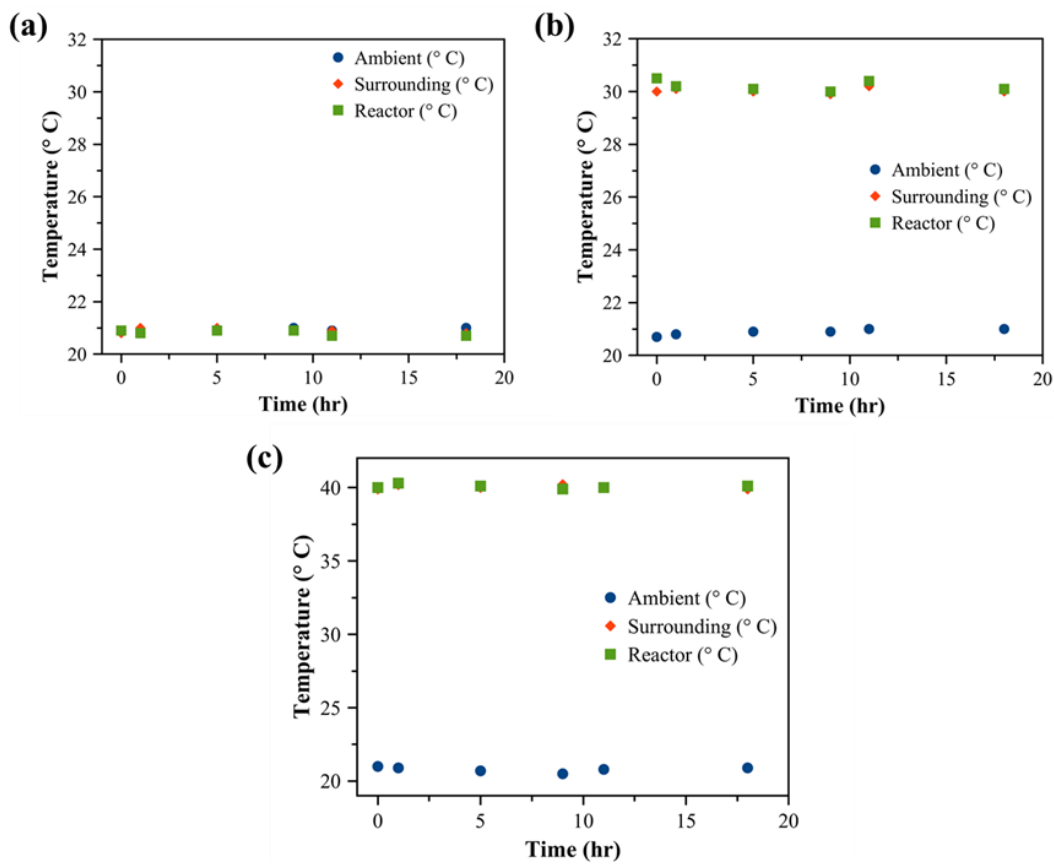

**Figure S1.** The temperature profile of methylene blue degradation in DMF under different conditions mentioned below **(a)** Dark 20 °C **(b)** Dark 30 °C **(c)** 30 °C

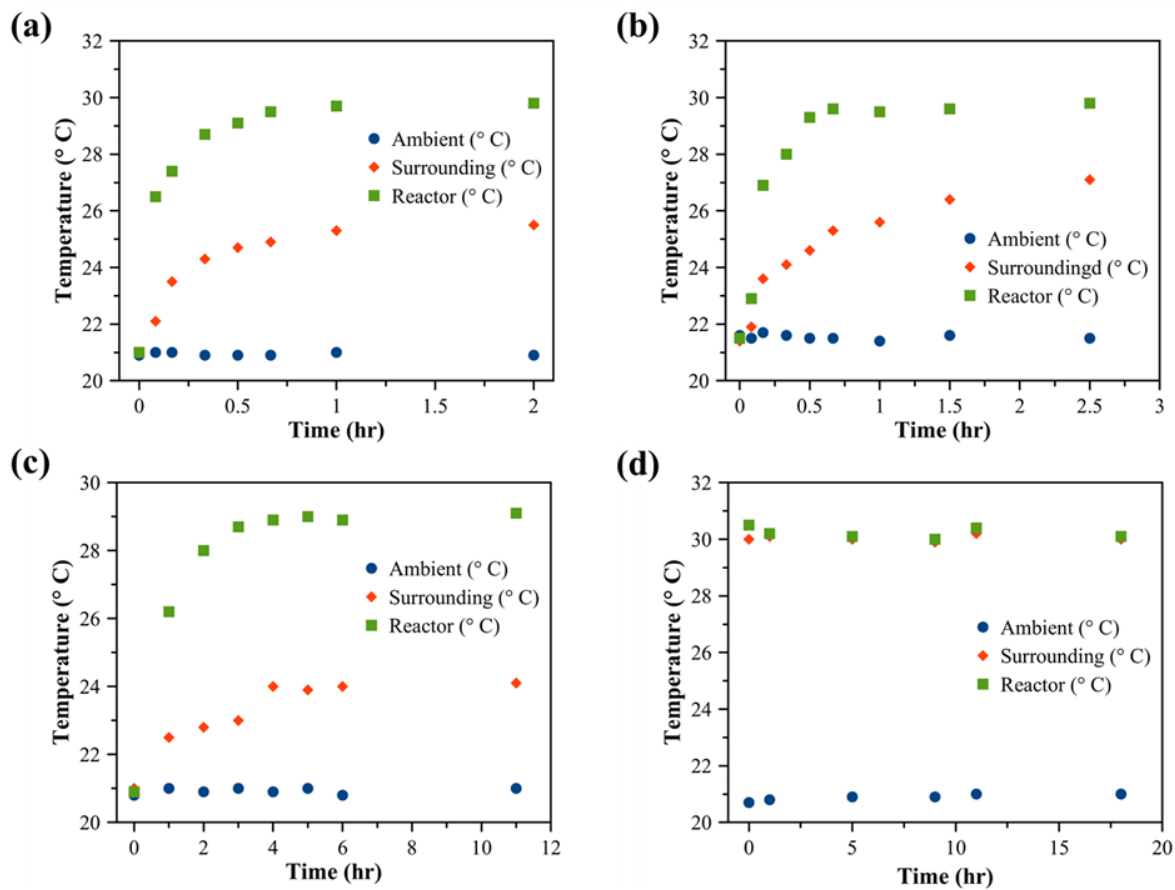

**Figure S2.** Temperature profile of methylene blue degradation in DMF under different conditions mentioned below (**a**) Red (**b**) Amber (**c**) Green illumination (**d**) Dark conditions at 30 °C

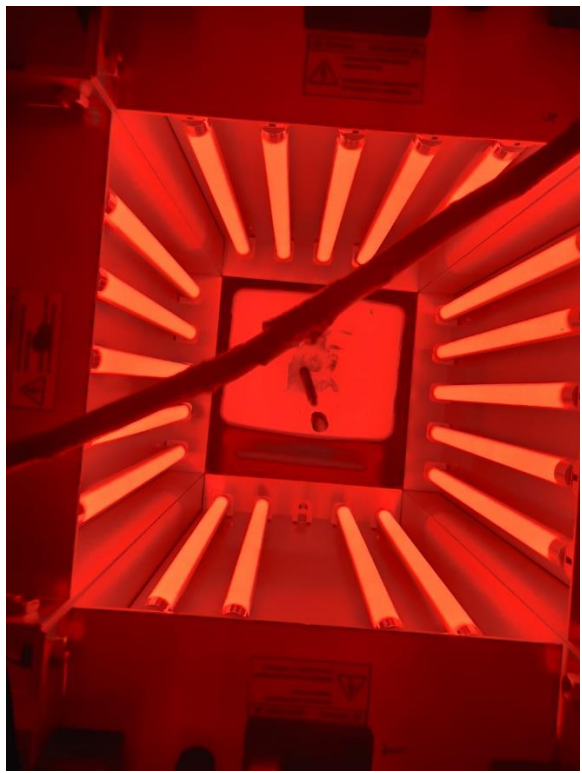

**Figure S2.** Luzchem Reactor set-up for performing reactions.